\documentclass[twocolumn]{aa}
\usepackage{natbib}
\bibpunct{(}{)}{;}{a}{}{,} % to follow the A\&A style
\usepackage{graphicx}
\usepackage{txfonts}

\newcommand{\bdtau}{\object{CFHT-BD-Tau~4}}

\newcommand{\micron}{$\mu$m}

\begin{document}

   \title{Grain growth and dust settling in a brown dwarf disk}

   \subtitle{Gemini/T-ReCS\thanks{
   {\small Based on observations obtained at the Gemini Observatory, which is operated 
   by the Association of Universities for Research in Astronomy, Inc., under a 
   cooperative agreement with the NSF on behalf of the Gemini partnership: 
   the National Science Foundation (United States), the Particle Physics and 
   Astronomy Research Council (United Kingdom), the National Research Council (
   Canada), CONICYT (Chile), the Australian Research Council (Australia), 
   CNPq (Brazil), and CONICET (Argentina).}} observations of CFHT-BD-Tau 4}

   \author{D. Apai
          \inst{1}
          \and
          I. Pascucci\inst{1}
	  \and
	  M. F. Sterzik\inst{2}
	  \and
	  N. van der Bliek\inst{3}
	  \and
	  J. Bouwman\inst{1}
	  \and
	  C. P. Dullemond\inst{4}
	  \and
	  Th. Henning\inst{1}
	  }

   \offprints{D. Apai, apai@as.arizona.edu}

   \institute{Max Planck Institute for Astronomy, K\"onigstuhl 17,
   D-69117 Heidelberg, Germany
   \and
     European Southern Observatory, Casilla 19001, Santiago 19, Chile
   \and
    {Cerro Tololo Inter-American Observatory, Casilla 603, La Serena, Chile}
        \and
    Max Planck Institute for Astrophysics, PO Box 1317, D-85741 Garching, Germany
             }

   \date{Received June 6, 2004; accepted June 24, 2004}

   \abstract{We present accurate mid-infrared observations of the disk
   around the young, bona-fide brown dwarf CFHT-BD-Tau 4. 
   We report GEMINI/T-ReCS measurements in the 7.9, 10.4 and 12.3 
   \micron{} filters, from which we infer the presence of 
   a prominent, broad silicate emission feature. 
   The shape of the silicate feature is dominated by emission from 
   2~\micron{} amorphous olivine grains. 
   Such grains, being an order of magnitude larger than those in the
   interstellar medium, are a first proof of dust processing and grain
   growth in disks around brown dwarfs.
    The object's spectral energy distribution is below
   the prediction of the classical flared disk model but higher than 
   that of the two-layer flat disk.
   A good match can be achieved by using an intermediate disk model with
   strongly reduced but non-zero flaring. Grain growth and dust settling
   processes 
   provide  a natural explanation for this disk geometry and we argue
   that such intermediate flaring might explain the observations of several other 
   brown dwarf disks as well. 
   }
   \maketitle

\section{Introduction}

Circumstellar disks play an important role in the formation and early evolution
of stars (see, e.g. \citealt{1987ARA&A..25...23S}), as well as
in the subsequent formation of planetary systems (see, e.g. \citealt{1993ARA&A..31..129L}). 
Similarly to young stars, infrared excess emission was recently identified  from brown dwarfs 
\citep{2000A&A...359..269C,2001ApJ...558L..51M,2003ApJ...585..372L,2003AJ....126.1515J}.
The interpretation of this excess emission as a proof of substantial amount of 
dust confined into a disk, was strongly supported by millimetre-wavelength observations 
estimating
the total mass of two brown dwarf disks being 0.4-6 M$_\mathrm{Jup}$ \citep{2003ApJ...593L..57K}.

The detection of brown dwarf disks sets at least two very important questions: 
{\em How does the disk structure compare to that of the T Tauri disks? 
Can brown dwarf disks potentially form planetary systems?}

Recent studies investigated the mid-infrared photometry from about 15 brown dwarf
disks aiming to determine the disk structure (see, e.g.
\citealt{2001A&A...376L..22N,2002ApJ...573L.115A,2003ApJ...590L.111P,2004ApJ...609L..33M,2004MNRAS.351..607W}). 
The two
basic models used  are the flat (\citealt{1988ApJ...326..865A}) 
and flared (opening angle increasing with the radius,
\citealt{1997ApJ...490..368C}) disks. Due to their geometry the flared disks
absorb a larger fraction of the stellar light. This additional energy is
re-emitted at mid- and far-infrared wavelengths leading to a far-infrared bump
often observed in the spectral energy distribution of T Tauri and Herbig
Ae disks. 

The flat and flared disk model predictions  deviate strongly longwards of
15~\micron{}. In spite of the lack of observations in this regime, 
there is mounting evidence that both disk structures can be found in brown dwarf
disks.

The similarity of disk structures and stellar mass/disk mass ratios between young
stars and brown dwarfs emphasizes the exciting question whether planet formation
can take place in brown dwarf disks. The major steps leading to 
planet formation start by grain growth and dust settling. 
While the early phases of such
grain coagulation has been recently observed in the disk atmosphere  (e.g.,
\citealt{2003A&A...412L..43P,2003A&A...400L..21V,2003A&A...409L..25M}) and
in the disk midplane (e.g.,
\citealt{2002ApJ...568.1008C,2003A&A...403..323T,2004A&A...416..179N}) of 
young low- and intermediate-mass stars,
 no such observation could be performed for the much fainter brown dwarf disks.

In this letter we focus on the disk around the spectroscopically 
confirmed, non-accreting brown dwarf \bdtau{} (\citealt{2001ApJ...561L.195M,2003ApJ...592..282J}, 
R.A. 04$^{\rm h}$ 39$^{\rm m}$ 47.3$^{\rm s}$ Dec.  +26$^{\rm h}$ 01\arcmin{}
39\arcsec{} J2000) being
the best-understood such object with measurements in the near-infrared
\citep{2003ApJ...585..372L, 2001ApJ...561L.195M}, mid-infrared ISOCAM \citep{2003ApJ...590L.111P} and sub-millimetre regimes 
\citep{2003ApJ...593L..57K}.  Based on these photometry \citet{2003ApJ...590L.111P} 
carried out the most detailed and systematic analysis of disk models. They
excluded all but three models as possibilities: the single- or double-layered 
flat disk or the flared disk with inner rim. 
These two models differ in the far-infrared and in the presence
of the 9.7~\micron{} silicate feature.
As a continuation of this effort we present new, 
accurate mid-infrared photometry for \bdtau{} obtained by
the recently installed T-ReCS camera. The filter bandpasses probe the region of
the 9.7~\micron{} silicate emission feature. Modeling the shape of this
feature allows us to  
determine the disk structure and test the dust grain properties.

\section{Observations, Data Reduction and Results}

The observations have been carried out on  Jan 2, 2004 using 
the T-ReCS mid-infrared detector mounted on the Gemini South 8m-telescope in service mode. The plate scale was
0.09\arcsec{}/pixel with a field of view of about 29\arcsec{}$\times$22\arcsec{}. 
Three filters were used with central wavelengths (50\% transmission wavelengths) of 
7.9~\micron{} (7.39-8.08~\micron), 10.38~\micron{} (9.87-10.89~\micron) and
12.33~\micron{} (11.74-12.92~\micron).

In order to eliminate the high thermal background the
usual chopping/nodding technique was used with throws of
10\arcsec.  The total on-source integration times
were 12, 7 and 10 minutes for the filters 7.9, 10.4 and 12.3~\micron. 
The data reduction has been carried out by using self-developed IDL routines. Each chopping pair
has been inspected manually to ensure the exclusion of the frames suffering from
strong detector signatures, such as sinusoidally modulated background noise. 
During this procedure we rejected about 10\% of the chopping pairs. Because the 
object is undetected in the individual chopping pairs, 
our manual selection does not bias towards stronger or fainter source flux.
After the manual inspection the high-quality
frames  have been  averaged. 

The \bdtau{} observations suffered from imperfect tracking, which resulted in 
somewhat blurred images. The relatively bright source, however, could be
identified also on subgroups of the images, allowing us to 
track and correct its position during the exposure. 
This correction proved that \bdtau{} is neither
extended nor elongated on the T-ReCS images and allowed more accurate
photometry.

The flux calibration is based on the calibrator HD~92305 ($\gamma$
Cha), which was observed several times during the night. In order to derive accurate flux densities
for the calibrator we integrated its spectra as given by \citet{1999AJ....117.1864C} within
the 50\% transmission levels of the T-ReCS filter set.
The flux measurements have been carried out by using aperture photometry with the IDL-adaptation of the
DAOPHOT routine \citep{1987PASP...99..191S}. 
To  increase the accuracy of our
photometry, we tested different aperture annuli. We found that the fluxes in the final 
two negative and one positive beams of the given source show 
consistent, slowly changing values for aperture radii between 0.9\arcsec --
1.35\arcsec. 
The fluxes obtained in this aperture range with 1-pixel steps 
have been used to determine the count rates from the source. 
These count rates have been converted to absolute flux densities
by comparing them to the mean of three count rate measurements of the standard star obtained by the corresponding
aperture sizes. 

% RESULTS
The mid-infrared emission of this brown dwarf disk is unresolved at a spatial
resolution of 0.3\arcsec{}. Our images prove
that no adjacent mid-infrared object could have influenced the previous
coarse-resolution ISOCAM photometry. 

We find the following flux densities for \bdtau{}: $F_{7.9}=42 \pm$10 mJy,
$F_{10.4}= 51.4\pm$2.6 mJy and $F_{12.3}= 47.1\pm$2.4 mJy. 
The final fluxes are the mean values over the aperture range, 
while the photometric error bars given are the maximum deviations from the mean value inside the 
aperture range combined with the calibration uncertainties.
The large error bar on the $F_{7.9}$ measurement reflects the low atmospheric transparency and the
short integration times.
The  excess emission at 10.4~\micron{} filter is undoubtly marking 
a prominent 9.7~\micron{} silicate feature.

\section{Models and Interpretation}

In this section we compare the new T-ReCS observations 
to model predictions  aiming to explore the dust properties and disk geometry. 
First, we will use the mid-infrared data to constrain the most dominant dust
species. Then, we will apply this information to a set of
self-consistent disk models with different geometries.

\subsection{Dominant Dust Species}
\label{Decomposition}
We model the dust composition by simply decomposing the observed
mid-infrared (T-ReCS and ISOCAM) flux densities into the sum of a 
black body and a modified black body, representing
the continuum emission and the optically thin component responsible for
the emission feature. A similar approach has been used in  
\citet{2001A&A...375..950B} and \citet{2003A&A...409L..25M}, and for the details
of the fitting procedure we refer to those articles.  The weighting factors of the two 
components and the common temperature
have been fitted by using the Levenberg-Marquardt minimization.
By introducing different optical constants to the optically thin component 
we tested dust grains of diverse types and sizes as possible dominant
species. We note, that the pure black body component can also represent
dust grains without spectral features in the mid-infrared regimes, such as carbonaceous
materials. The weighting factor of the modified black body is proportional
to the mass in the optically thin regime.

In our model we used silicates with  olivine composition
(FeMgSiO$_4$, \citealt{1995A&A...300..503D}), 
the most abundant interstellar silicate species, as well as amorphous silica
\citep{1960PhRv..121.1324S}.
For both species we fitted the observation using spherical grains with radii between 0.1 and 100~\micron.    
Based on the $\chi^2$ comparison we found, that the 2.0~\micron{}-sized
olivine  grains provide the best fit to the observations. 
The best-fit temperature was 490 K  and the
total dust mass in the optically thin regime equaled to 0.1 lunar mass.

In order to put a qualitative constraint on the mass ratio of the
large/small grains, we included a second
optically thin component. We find, that the lower 
limit for the mass ratio of the 2.0~\micron{} grains to the 0.1~\micron{}
ones in the optically thin disk regime is $m_{2.0}/m_{0.1} > 6$.

\subsection{Self-consistent disk models}

The fitting procedure described above has no self-consistent
physical basis, but it provides a good indication for the dominant dust species. 
We confirm this information by using  self-consistent disk
models described in \citet{2001ApJ...560..957D}, which are improved versions of 
models by \citet{1997ApJ...490..368C}. In these models the radiation field of the central object together with the
dust properties determine the disk flaring angle.

In \citet{2003ApJ...590L.111P} we already applied these models 
for \bdtau{} and compared several disk geometries. We
excluded all but three models:  single-layer flat disk, two-layer flat disk 
and flared disk with inner rim ({\em m4}, {\em  m5} and {\em m9} in \citealt{2003ApJ...590L.111P}). 
The main parameters of those and the following models are:
brown dwarf luminosity $L_{BD}=0.1~\rm L_\odot$, mass $M_{BD}=0.075$~M$_\odot$,
 temperature $T_{eff}=2800$~K; the dust opacities are those of
0.1\micron{} astronomical silicates \citep{1984ApJ...285...89D}. The outer radius was set to 
50~AU, but our observations do not constrain this parameter. 
The presence of the 9.7~\micron{} silicate feature arises from an optically thin
disk layer (e.g., disk atmosphere). As the single-layer flat disk model lacks
any optically thin component, this model can be immediately excluded. 
 When comparing the {\em m9} flared and {\em m5} flat disk models, we find that none
of them matches well the new observations as long as the 0.1~\micron{}
astronomical silicates are used as dust species.
Thus, successfully fitting the detailed mid-infrared data requires the identification
of the dust species and its typical grain size in addition to the disk
structure. 

Here we use the results of Sect.~\ref{Decomposition},  namely, our guess of
2.0~\micron{} olivine grains as the dominant dust species. 
When we apply 2.0~\micron{}-sized amorphous olivine
grains in the disk models, we find that the two-layer flat disk model ({\em m5}) 
reproduces the observations better than the flared disk model. 
The best fit, however, is reached by applying a disk model
intermediate between the flat and flared disk models: as discussed in the
Sect.~\ref{Discussion}  we artificially reduced the disk height by a factor of 4
to account for dust settling effects.

\begin{figure}
   \centering
   \includegraphics[width=9cm]{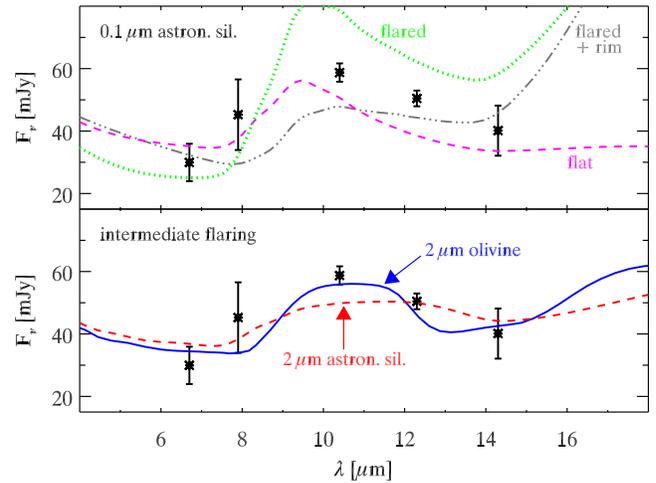}
      \caption{Flat, flared and intermediate
  disk models compared to the observed
       spectral energy distribution of \bdtau. 
                 { The upper panel has been
       calculated by using 0.1 astronomical silicate grains, while the lower panel
       applies 2.0~\micron{} olivine and astronomical silicate grains.  All model
       parameters are identical to those in \cite{2003ApJ...590L.111P}, except
       for the reduced flaring parameter of the intermediate model. 
       The 850~\micron{} and 1.3~mm observations are not shown here but fitted
       well, similarly to \citet{2003ApJ...590L.111P}.  
       The best fit is achieved by the intermediately flaring model in
       combination with 2\micron{} amorphous olivine grains.       }
       }
         \label{FlaredvsFlat}
   \end{figure}

As an illustration of the influence of the main parameters and the stability of
our results in Fig.~\ref{FlaredvsFlat} we show the best-fit models with flat and flared geometries in
combination with interstellar-like 0.1~\micron{} astronomical silicate grains,
as well as the intermediately flaring disk model and 2~\micron{} grains.

Throughout the modeling procedure we also varied the  inclination $i$,  the inner
disk radius $R_{in}$, 
the  exponent $\alpha$ of the surface density power-law $r^\alpha$, the presence
of an inner rim and self-shadowing effects, the latter two being self-consistently
calculated. The best solutions are found around the values of $i \simeq0\degr$,
$\alpha=-1.9$,
$R_{\rm in}=3.5~R_{\rm BD}$, and no inner rim.

Independently of the grain size estimate obtained by decomposing the emission
feature,  we also tried to fit the observations by using 
different dust species and geometries. However, no other configuration than a disk
with reduced flaring and moderate-sized (2~\micron{}) olivine grains 
could reproduce the measurements.

\section{Discussion}
\label{Discussion}
\subsection{Disk structure and Dust settling}

The sparsely sampled spectral energy distributions of the brown dwarf disks
observed until now have been generally modeled by either flat or flared disk
geometries. The flared disk models are based on the assumption of efficient 
mixing of gas and dust everywhere in the disk \citep{1997ApJ...490..368C}. In this picture
the dust particles are kept above the disk midplane by the turbulent gas, 
which in turn is thermally coupled to the dust grains absorbing the incident stellar
radiation. Grain growth, however, reduces the dust-gas coupling \citep{1995Icar..114..237D,2004astro.ph..5226D}
 and accelerates the sinking of dust particles 
toward the disk midplane. Thus, the {\em dust} disk converges towards the 
flat disk geometry. Reduced disk scale heights have been argued for in the case
of several disks around low- and intermediate-mass stars (see, e.g.
\citealt{2001ApJ...547.1077C}).

Our modeling yields two major results: 1, the optically thin disk regime
is dominated by at least intermediate-sized ($\sim$2\micron) dust grains; 2, 
the dust disk's scale height is strongly reduced. 

These results are fully consistent with the evolution of disks through grain
growth and subsequent settling, as outlined above. Up to now, this process has
been only rarely observed in circumstellar disks and never in a disk of a brown
dwarf. The multi-wavelength observations of the relatively bright disk of \bdtau{} provide
the first evidence that the similarity between young stars and brown dwarfs
extends to circumstellar dust processing and the subsequent evolution of the
disks. 

We note here, that the picture of grain growth {\em and} dust settling might provide
a natural explanation not only for \bdtau{} but for several other brown dwarf
disks with flat disks, such as \object{Cha H$\alpha$2} \citep{2002ApJ...573L.115A}, 
\object{GY5} \citep{2004ApJ...609L..33M} and further 9 objects from
\citep{2001A&A...376L..22N,2002A&A...393..597N}.

{ In Sect.~\ref{Decomposition} we estimated the mass in the optically thin
regime  to be $\sim$0.1 lunar mass. The same method when applied to the sample
of
Herbig Ae stars presented in \citet{2001A&A...375..950B} results in  masses of
$\sim100$ lunar masses, about a thousand times larger than
that of \bdtau. If these disks would be optically thin, their mid-infrared
emission ratio would be similar to their mass ratio, i.e. $\sim 50$. The larger
difference, however, is naturally explained if the emission originates from the
atmosphere of an optically thick disk. In this case, the total emission scales
with the surface area contributing to the 9.7~\micron{} silicate feature: The
radial dust temperature distribution in the disk atmosphere approximately scales
as  $T(r) \propto T_* \times \large(r\large)^{-{2 \over 5}}$, where $r$ is the radius,
$T_*$ is the stellar temperature, assuming the absorption coefficient
$\kappa$ to be $\propto \nu$. Given the typical Herbig Ae stellar temperatures being
$\sim$4 times larger than the corresponding brown dwarf temperatures, any given temperature around a brown dwarf is reached at
$\sim30$ times smaller radii compared to the Herbig Ae stars. This translates to
a surface ratio of $\sim10^3$, the same order of magnitude as the
fitted  mass ratios.}

Our best-fit model --- similarly to those of
\citet{2001A&A...376L..22N}, \citet{2004ApJ...609L..33M}, \citet{2004MNRAS.351..607W} and
others --- require an inner disk radius of $R_{\rm in}\simeq 3.2~\rm R_{\rm BD}$. This value
is similar to the typical inner disk truncation radii deduced for T Tauri
stars \citep{1994ApJ...429..781S} { and might be defined through dust sublimation at
temperatures of $\sim1500$K.}

\subsection{Grain growth in a substellar disk}

Our observations demonstrate that the disk atmosphere of \bdtau{} is dominated by
2.0~\micron{} grains, an order of magnitude larger than the
average grains commonly found in the interstellar matter.
Further  modeling indicates that the grain growth is accompanied by dust
settling. These two processes are the initial, essential steps 
of planet formation \citep{1993ARA&A..31..129L}. 

Studies of coeval T Tauri disk systems showed that the
dust evolutionary stage is only a weak function of time (see, e.g.
\citealt{2003A&A...412L..43P,2003A&A...409L..25M}). On the other hand, 
the rate of dust processing is likely to provide a useful way to parameterize
the evolution of an accretion disk through protoplanetary disk into
a debris disk.
Comparing the derived $m_{2.0}/m_{0.1}$ ratio to those estimated for disks of 
Herbig Ae stars \citep{2001A&A...375..950B} provides some  insight into the
stage of the dust evolution in \bdtau.  The values range from
$m_{2.0}/m_{0.1} < 0.04$ for the unprocessed dust in the Galactic Centre up to $> 57$ for the 
evolved disk of \object{HD~100546}. 
 The $m_{2.0}/m_{0.1}>6$ ratio of \bdtau{} indicates that this
brown dwarf disk is close in dust evolution phase to the 5~Myr-old \object{HD~104237} \citep{2004ApJ...608..809G}.
The fact, that comets Halley and Hale-Bopp show  the 
presence of smaller grains ($m_{2.0}/m_{0.1} <$ 3), might suggest that they 
formed in a dust evolutionary phase -- but not necessarily in time -- {\em prior} to that of \bdtau.

{ We directly derive the feature strength from the observations following \citet{2003A&A...412L..43P} and
compare its $\sim$1.6 value with the T Tauri star samples of \citet{2003A&A...409L..25M} and
\citet{2003A&A...412L..43P}. This model-independent comparison again shows 
\bdtau{} among the disks with the most processed dust.}

The young age of \bdtau{} ($\sim$1~Myr, \citealt{2001ApJ...561L.195M}) stands
{ intuitively} in contrast to the
evolved stage of its dust disk. There are three factors which may resolve
this apparent contradiction: First, large differences in dust processing
timescales have been spotted among T Tauri and Herbig Ae stars, indicating
that dust processing can proceed at very different rates. 
Second, the over-luminosity of \bdtau{} may be explained by an unresolved
binary brown dwarf. Dynamical perturbations could then { influence 
 the dust coagulation and settling timescales}. Hints for this behavior have been found in T Tauri
disks \citep{2003A&A...409L..25M} and other brown dwarf disks \citep{2004A&A...submitted}.
Third, if it is a binary system, then the age of \bdtau{} is likely to be underestimated.

The only previous detection of silicate emission from a brown dwarf disk was
found for the $\rho$ Ophiucus object \object{GY310} \citep{2004ApJ...609L..33M}.
In that case the shape of the feature hints on small dust grains, but the data
quality do not allow any firm conclusion. Spectroscopic surveys
with the new Spitzer Space Telescope will be able to derive the accurate dust
composition for a large number of brown dwarf disks revealing how frequent
grain growth processes are. However, to follow the grain growth to scales
larger than a few micron, high-resolution (sub)millimetre observations are
necessary, possibly by using interferometric facilities such as the
Sub-millimeter Array or ALMA.

\section{Summary}

The main conclusions of this work are the following:
\begin{enumerate}
     \item Using the T-ReCS/Gemini instrument we detected thermal infrared
     emission at 7.9~\micron, 10.4 \micron{} and 12.3 \micron{} from the disk of \bdtau.
     \item The disk  displays a prominent silicate emission peak 
      which proves the existence of an optically thin disk layer.      
     \item We compare the shape of the silicate emission feature to 
      laboratory spectra emerging from different dust species and 
      grain sizes. Using two independent methods we find that 
      the emission feature is dominated by 2~\micron{} amorphous silicate
      (olivine) grains.             
        \item Simple semi-analytical models  explain the observations 
      by a two-layered flared disk with a reduced scale height. 
      This transitionary disk model between flat and flared geometries is 
      consistent with dust settling models.

       \item The dominance of 2~\micron{} grains and the reduced disk flaring
        prove that grain growth and settling occurred in the disk of the
	$\sim$1~Myr-old \bdtau{}. 
	Based on the derived mass ratio for the 2~\micron{} to 0.1~\micron{} grains
	the stage of dust processing in this brown dwarf disk is similar to 
	that of the well-known Herbig Ae star HD~104237.
	
\end{enumerate}     
In view of these results we conclude, that the dust grain coagulation and
dust settling, the initial steps of planet formation, take place in  
the disk of the brown dwarf \bdtau.

\bibliographystyle{aa}
\bibliography{lit.bib}

\end{document}